%% file: main.tex
\documentclass[conference]{IEEEtran}
\IEEEoverridecommandlockouts

\usepackage{cite}
\usepackage{amsmath,amssymb,amsfonts}
\usepackage{algorithmic}
\usepackage{graphicx}
\usepackage{textcomp}
\usepackage{xcolor}
\usepackage{verbatim}
\usepackage{caption}
\usepackage{subcaption}
\usepackage{tabto}
\usepackage{balance}
\usepackage[T1]{fontenc}
\usepackage[hidelinks]{hyperref}
\usepackage[a4paper, total={184mm,239mm}]{geometry}
\def\BibTeX{{\rm B\kern-.05em{\sc i\kern-.025em b}\kern-.08em
    T\kern-.1667em\lower.7ex\hbox{E}\kern-.125emX}}
\begin{document}

\newcommand{\hms}[1]{\textcolor{blue}{HMS: #1}}

\renewcommand{\sectionautorefname}{Section}

\newif\ifremark
\long\def\remark#1{
\ifremark%
        \begingroup%
        \dimen0=\columnwidth
        \advance\dimen0 by -1in%
        \setbox0=\hbox{\parbox[b]{\dimen0}{\protect\em #1}}
        \dimen1=\ht0\advance\dimen1 by 2pt%
        \dimen2=\dp0\advance\dimen2 by 2pt%
        \vskip 0.25pt%
        \hbox to \columnwidth{%
                \vrule height\dimen1 width 3pt depth\dimen2%
                \hss\copy0\hss%
                \vrule height\dimen1 width 3pt depth\dimen2%
        }%
        \endgroup%
\fi}

\remarkfalse

\title{Optimizing Energy Efficiency in \\ Subthreshold RISC-V Cores}

\author{Asbjørn Djupdal, Magnus Sj\"{a}lander, Magnus Jahre, Snorre Aunet, and Trond Ytterdal \\
\textit{Norwegian University of Science and Technology, NTNU}, Trondheim, Norway \\
Asbjorn.Djupdal | Magnus.Sjalander | Magnus.Jahre | Snorre.Aunet | Trond.Ytterdal@NTNU.no}

\maketitle

\begin{abstract}
Our goal in this paper is to understand how to maximize energy efficiency when designing standard-ISA processor cores for subthreshold operation.
We hence develop a custom subthreshold library and use it to synthesize the open-source RISC-V cores SERV, QERV, PicoRV32, Ibex, Rocket, and two variants of Vex, targeting a supply voltage of 300\,mV in a commercial 130\,nm process.
SERV, QERV, and PicoRV32 are multi-cycle architectures, while Ibex, Vex, and Rocket are pipelined architectures.

We find that SERV, QERV, PicoRV32, and Vex are Pareto optimal in one or more of performance, power, and area.
The 2-stage Vex (Vex-2) is the most energy efficient core overall, mainly because it uses fewer cycles per instruction than multi-cycle SERV, QERV, and PicoRV32 while retaining similar power consumption.
Pipelining increases core area, and we observe that for subthreshold operation, the longer wires of pipelined designs require adding buffers to maintain a cycle time that is low enough to achieve high energy efficiency.
These buffers limit the performance gains achievable by deeper pipelining because they result in cycle time no longer scaling proportionally with pipeline stages. The added buffers and the additional area required for pipelining logic however increase power consumption, and Vex-2 therefore provides similar performance and lower power consumption than the 5-stage cores Vex-5 and Rocket.
A key contribution of this paper is therefore to demonstrate that limited-depth pipelined RISC-V designs hit the sweet spot in balancing performance and power consumption when optimizing for energy efficiency in subthreshold operation.
\end{abstract}


\input{introduction}
\input{bowilib}
\input{cores}
\input{experiment}
\input{results}
\input{conclusion}

\section*{Acknowledgements}

We thank Olof Kindgren for providing us with early access to the
RISC-V QERV core implementation.
The work has been funded in part by the European Union’s Horizon 2020 research and innovation program through the  BOWI project (grant agreement number 873155).

\bibliographystyle{ieeetr}
\bibliography{refs}

\end{document}

%% file: introduction.tex
\section{Introduction}

Many emerging applications within important domains, such as healthcare and the Internet of Things (IoT), have extremely stringent energy requirements.
At the same time, it is typically not economically viable to create special-purpose ASICs for each type of application.
Developers are hence forced to use ultra-low-power (ULP) processors in these systems, but the active power consumption of most commercial ULP devices is still measured in milliwatts and hence falls short in many deployments. Biomedical applications~\cite{chen_injectable_2015}, for example, require power consumption at the level of microwatts or below.

It is well known that reducing the supply voltage into the subthreshold region yields energy-minimal operation for a given design~\cite{weak-inversion:JSSC1997, Tsiv}, but it is yet not clear how to design energy efficient microcontrollers for subthreshold operation.
Zhai et al.~\cite{subVt-mcu:TVLSI2009} evaluate energy-efficient subthreshold processor designs, but a key part of their study is the design of the instruction set architecture (ISA) itself.
While designing a new ISA enables optimizations for the target application domain, e.g., specialized registers and a low-precision arithmetic unit (ALU), it also incurs two significant drawbacks.
First, special-purpose ISAs lack a software ecosystem, but Arm and RISC-V have demonstrated that a thriving software ecosystem is a key enabler of commercial success. 
Second, special-purpose ISAs are potentially over-constrained and therefore only efficient for a few applications. 
This significantly reduces the reuse benefits that make general-purpose microcontrollers cost-effective.

In contrast, basing the processor design on a standard ISA, such as Arm or RISC-V, enables tapping into a software ecosystem of widely used and industry-proven compilers, libraries, and operating systems, and such ISAs are designed for flexibility, which makes them efficient for a wide range of applications.
Examples of prior work on subthreshold microcontrollers with a standard ISA, e.g., Zscale~\cite{zscale:ESSCIRC2018} (RISC-V), SleepRunner~\cite{sleeprunner:JSSC2021} (ARM) and Bottle Rocket~\cite{10409181} (RISC-V), focus on evaluating a single microcontroller and therefore do not shed light on the key trade-offs of subthreshold microcontroller design.
Furthermore, each implementation is evaluated in a different process technology, making it challenging to extract generic conclusions from the various works.
Conti et al.~\cite{schiavone:PATMOS2017} evaluates three RISC-V microcontroller designs, but use a standard cell library with a nominal supply voltage of 1.2\,V and only report results for supply voltages down to 0.8\,V (which is far above the subthreshold region).

Our goal in this paper is to fill this knowledge gap in the design of subthreshold standard-ISA microcontrollers.
More specifically, we want to understand how to design the microarchitecture of standard-ISA microcontrollers to maximize energy efficiency when targeting subthreshold operation.
Since commercial subthreshold standard cell libraries are not available\footnote{Commercial subthreshold microcontrollers exist, e.g., Ambiq's SPOT-enabled devices~\cite{ambiq-spot}, but, to the best of our knowledge, no commercial subthreshold cell library is available.}, we designed our own library for a commercially available 130\,nm process technology node (\autoref{sec:bowilib}); this is a commonly used technology node for microcontrollers.
We have characterized the cell library and verified functional correctness using simulations at supply voltages between 250 and 400\,mV.

We use our subthreshold library to evaluate several open-source RISC-V cores that exhibit a wide range of microarchitectural features (\autoref{sec:cores}).
SERV~\cite{serv:GITHUB}, QERV~\cite{serv:GITHUB}, and PicoRV32~\cite{picorv32:GITHUB} represent multi-cycle architectures, meaning that (1)~a single instruction is executed at a time, and (2)~each instruction takes multiple cycles to execute.
SERV implements a bit-serial ALU, which yields low area at the cost of increased instruction latency, while the QERV operates on four bits at a time, which increases area but significantly lowers latency compared to SERV.
PicoRV32 is a more conventional multi-cycle core with a standard bit-parallel ALU.
Ibex~\cite{ibex:GITHUB}, Vex~\cite{vexriscv:GITHUB}, and Rocket~\cite{rocket} represent pipelined architectures, which means that multiple instructions overlap and are executed in parallel.
Ibex and Vex-2\footnote{The pipeline depth of the Vex core is configurable and Vex-2 represents a 2-stage pipeline while Vex-5 represents a 5-stage pipeline.} represent shallow pipelines with two stages while the Rocket and Vex-5 represent deeper pipelines with five stages.
All cores implement the RV32E ISA, which is the minimal RISC-V ISA with only 16 registers and no hardware support for multiplication, division, or floating-point.
Each core is taken through synthesis, place and route, and parasitic extraction using an industrial-grade tool flow, 
and evaluated in the subthreshold region at 300\,mV, which is close to the energy-optimal supply voltage for our cell library (\autoref{sec:experimental-setup}).

We use eight benchmarks from MachSuite~\cite{machsuite:IISWC2014} to evaluate performance and energy efficiency; our energy efficiency metric is energy per instruction (EPI).
We find that Vex-2 yields the lowest EPI due to two main reasons:
\textit{First}, multi-cycle SERV, QERV, and PicoRV32 use (many) more cycles per instruction (CPI) compared to Vex-2.
Therefore, their low power consumption does not compensate for their long runtime.
\textit{Second}, for the deeply pipelined Vex-5 and Rocket, the additional buffers inserted during synthesis to enable subthreshold operation means that the clock cycle time does not decrease enough compared to Vex-2 to compensate for the increase in CPI caused by the deeper pipeline.
They hence yield similar performance to Vex-2 but with increased power consumption.
We further study EPI in the context of runtime, power, and area constraints and find that other cores than Vex-2 can be Pareto optimal.
For example, Vex-5 yields (slightly) lower runtime than Vex-2, and the multi-cycle SERV and QERV are Pareto optimal when power or area is (significantly) constrained.

%% file: bowilib.tex
\section{Custom Subthreshold Library}
\label{sec:bowilib}

This paper uses a commercially available 130\,nm process for which the nominal supply voltage is 1.2\,V and the absolute values of the threshold voltages are around 350\,mV.  A standard cell library in this 130 nm technology for a supply voltage of 1.2 V is available, but not optimized, nor characterized for supply voltages below 1.08\,V.
Lack of characterization means that the standard library can not be used by EDA tools below 1.08\,V.
Even if the library was characterized for other voltages, the lack of optimization for subthreshold operation likely makes the library perform suboptimally at ultra-low voltages~\cite{937856}.

We have developed a custom cell library to perform the experiments in this paper.
This section goes through the design of this library and explains design choices affecting subthreshold operation, in addition to details on verification of the library.
\subsection{Cell Selection}

\begin{table}[t]
  \centering
  \caption{Custom cell library.}
  \begin{tabular}[tp]{l|l}
    Type & Cell \\
    \hline
    \hline
    Basic gates & INV\_X\{1,2,5,10,20,40,80,160,320\}, \\
                & NAND\_X\{1,2,5,10\}, NOR\_X\{1,5,10\}, \\
                & XOR\_X5, XNOR\_X5, AOI\_X5, OAI\_X5 \\

    Buffers & BUF\_X\{2,5,10,20,40,80,160,320\}, DLY\_X\{4,8\} \\

    Latches & DLATCH\_X5 \\

    Flipflops & DFF\_X5, DFF\_SET\_X10, DFF\_RESET\_X10 \\

    Complex cells & FA\_X5, MUX\_X2 \\
  \end{tabular}
  \label{tab:bowilib}
\end{table}

\autoref{tab:bowilib} describes our custom cell library.
While more cells typically improve the efficiency of the synthesized circuit, subthreshold cell libraries with relatively few cells are not uncommon,
e.g., only four basic cells were used in~\cite{7062968}. A likely reason is that adding more cells incurs an engineering overhead.
We therefore include some additional commonly used cell types to improve efficiency, but keep the overall number of cells low to reduce engineering overhead.

The conventional 130\,nm standard cell library contains many logic gates with high fan-in.
Noise margins are worse in subthreshold~\cite{bol2011robust} because it gets more difficult to achieve logic gate outputs close to either $V_{DD}$ or $V_{SS}$.
The transistor can be viewed as a voltage-controlled resistor, and in subthreshold the difference in resistance between a (somewhat) conducting state and a (mostly) non-conducting state is lower than for conventional supply voltages.
Combined with high loads on logic gate outputs, the output voltages may get degraded or take long to settle.
This makes it important to keep a low number of transistors in series between a logic gate output and either $V_{DD}$ or $V_{SS}$~\cite{calhoun2005modeling}, dictating a low fan-in.  We limit the logic gates in our library to a maximum of two transistors separating cell output from either $V_{DD}$ or $V_{SS}$, resulting in mostly 2-input gates.
The FA cell is a conventional 28T full adder~\cite{full-adder:ISSCC1990} and has three transistors in series, but not on the output which is driven by an inverter that restores internal signals.

\subsection{Transistor Sizing}

Logic gates, if designed such that all transistors have the same size, will likely exhibit an imbalance with respect to rise and fall times.  
This is unfortunate as it will negatively affect both switching speeds and power.  A key target when choosing the transistor sizes of a cell is, therefore, to aim for a balance between the nMOS and pMOS networks for the relevant supply voltages and operating temperatures.

We choose transistor sizes based on simulations of toggle times.  A simulation setup is created for each logic gate such that the gate inputs can be toggled independently.
The latency from an input toggles to the output passes $V_{dd}/2$ is then simulated.  This is done for all input transitions that lead to a rising ($0 \rightarrow 1$) or falling ($1 \rightarrow 0$) output transition, and the average rise latency $R$ and the average fall latency $F$ are calculated.
We then devised a metric, called \emph{imbalance}, to quantify the extent to which a logic gate exhibits imbalance in rise and fall latencies: 
$i = \frac{\lvert R - F \rvert}{R + F}$.
The imbalance $i$ is a number between 0 and 1 where a logic gate with perfectly balanced rise and fall times yields $i$ equal to 0 while a completely imbalanced logic gate yields $i$ equal to 1. Minimizing $i$ is hence the goal when sizing the transistor gates. 
Load affects the degree of imbalance, and we therefore let all simulated gates drive an inverter.

Minimum sized transistor gates should be avoided in subthreshold circuits.
Local random mismatch (LRM) causes transistors that should ideally have similar electrical properties to behave differently due to limitations in the fabrication processes.
Pelgroms law~\cite{pelgrom} states that the variance of the threshold voltage due to mismatch is inversely proportional to the transistor gate areas $width \times length$ ($WL$).
There are exponential dependencies governing the current levels in subthreshold operation, and the threshold voltages, therefore, have large effects on subthreshold operation.  This makes it important to minimize the effects of mismatch on the threshold voltage.  
The default nMOS gate area ($WL$) is, therefore, chosen to be five times the minimum size.  This reduces the threshold voltage variance to 20\% compared to the minimum sized transistor.  We sometimes deviate from this guideline to achieve the necessary drive strength and performance.  

The pMOS gate area is then determined based on a transient simulation sweeping the pMOS area.  This is done for all logic gates. The pMOS area minimizing the imbalance metric $i$ is chosen.  More complex cells are constructed from logic gates and do not need additional balancing. 
When choosing transistor gate area, the length is set to minimum size and only the width is customized for the cell. 

\subsection{Cell Library Layout and Verification}
\label{sec:cell_lib_verification}

\noindent
\textbf{Layout.}
The layouts of the cells are created in Cadence Virtuoso~\cite{virtuoso}.  
Post-layout simulations with parasitic extraction is performed to verify that the cells are functionally correct and perform as expected over the range of voltages and temperatures the library is expected to work with.
Simulations were performed with Cadence Spectre~\cite{spectre} and parasitic extraction with Siemens Calibre~\cite{calibre}.  After layout, the library was characterized with Cadence Liberate~\cite{liberate}.

 \begin{figure}[t]
   \centering
     \begin{subfigure}[b]{0.49\columnwidth}
       \centering
       \includegraphics[width=\columnwidth]{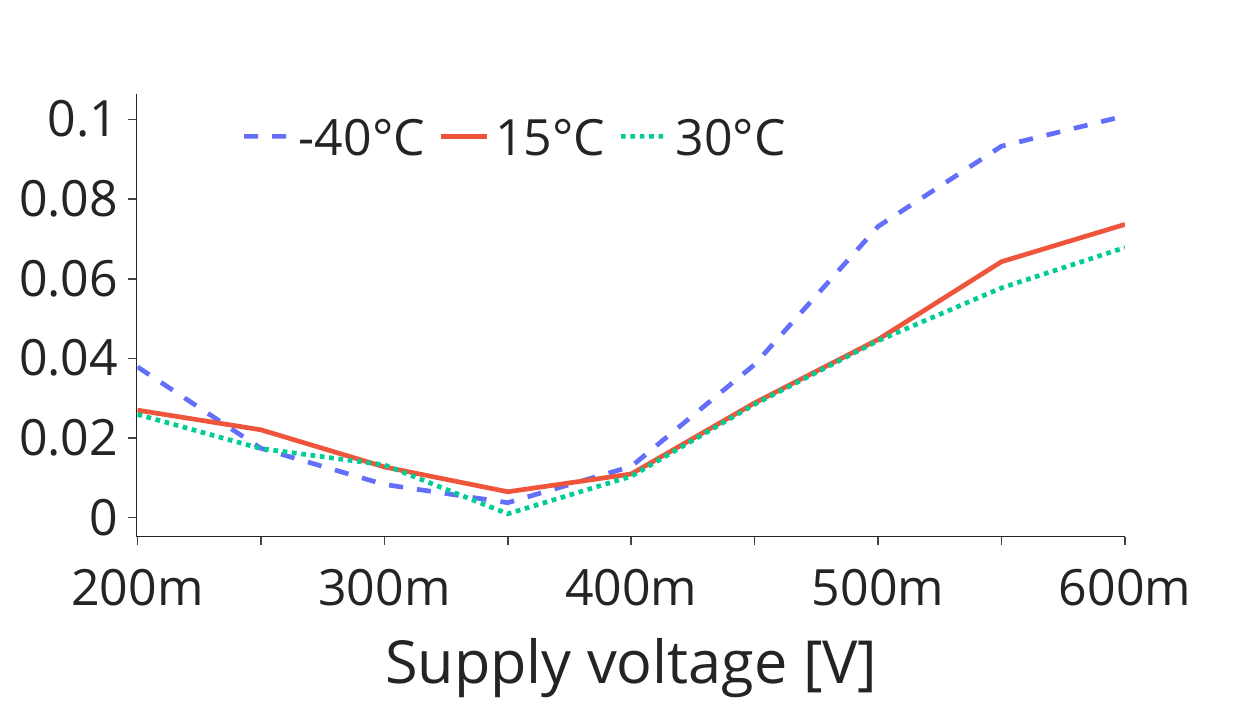}
       \caption{INV X5}
       \label{fig:balance_iv}
     \end{subfigure}
     \hfill
     \begin{subfigure}[b]{0.49\columnwidth}
       \centering
       \includegraphics[width=\columnwidth]{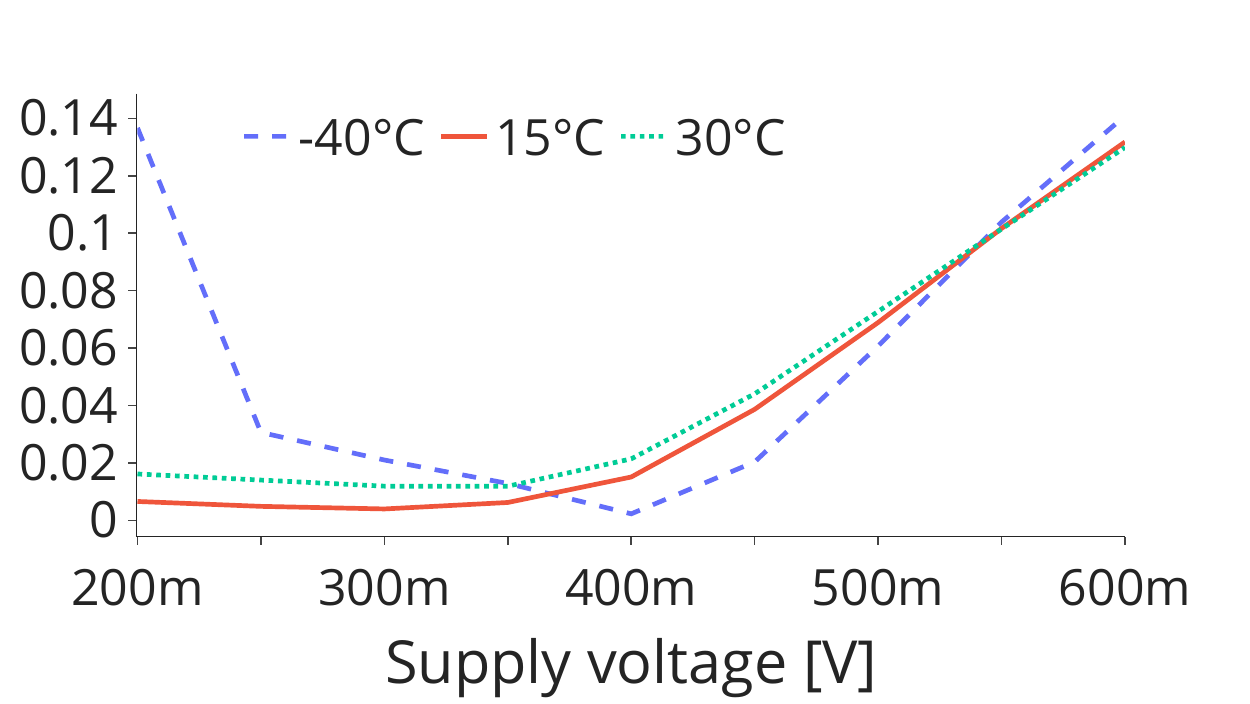}
       \caption{NAND X5}
       \label{fig:balance_nand}
     \end{subfigure}
     \caption{Imbalance $i$ versus voltage for across temperatures.}
     \label{fig:balance}
\end{figure}

\autoref{fig:balance} shows an example of a post layout simulation of the inverter (INV) and NAND cells of size X5.
Imbalance $i$ is depicted against supply voltage for three different temperatures (-40\textdegree C, 15\textdegree C, and 30\textdegree C).
These cells are reasonably balanced by staying below 3\% for all temperatures and voltages between 250\,mV and 400\,mV.  They should, therefore, be well suited for the temperature and voltage range the library is targeting.  

\noindent
\textbf{Verification.}
The library is designed to work with a supply voltage between 250\,mV and 400\,mV, and local random mismatch (LRM) should, therefore, not cause functional defects.
To verify this, several designs were implemented with a full physical layout and then simulated with a transistor-level analog simulation including extracted parasitics.
The analog simulations were performed at the lowest supply voltage of 250\,mV, and with Monte Carlo simulations investigating LRM.
The largest design was PicoRV32~\cite{picorv32:GITHUB}, which is one of the RISC-V cores evaluated later in this paper.  As the core does not contain memory, and also requires some control signals set externally, a Verilog testbench had to be implemented that contains the memory and controls the interface to the PicoRV32 core.
The instruction memory is filled with a small test program.  As these simulations are very time-consuming, only a very small test program containing 18 instructions could be simulated to completion within reasonable time.
Cadence Xcelium~\cite{xcelium} was used to co-simulate the digital Verilog testbench and the analog PicoRV32 core.
The core was simulated until successful completion of the 18-instruction test program. 
No failures were detected with any of the tested designs, including PicoRV32.  

%% file: cores.tex
\section{The RISC-V Cores}
\label{sec:cores}

\begin{figure}[t]
    \centering
    \begin{subfigure}[b]{\columnwidth}
         \centering
         \includegraphics[width=0.8\linewidth]{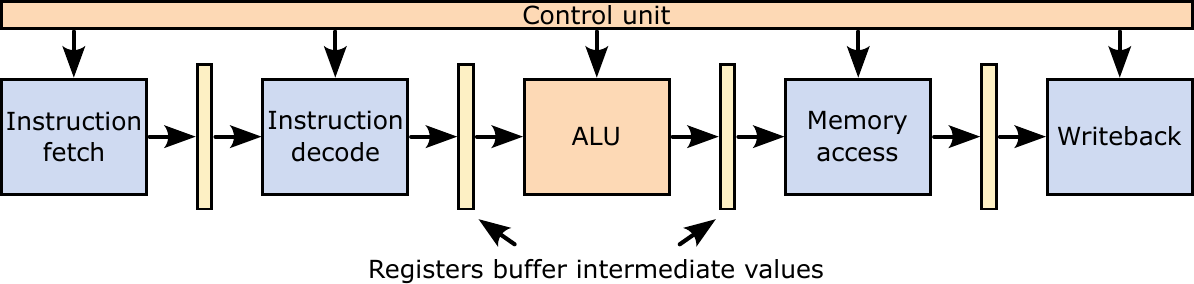}
         \caption{Multi-cycle architecture.}
         \label{fig:microarch-multi-cycle}
     \end{subfigure}
     
     \begin{subfigure}[b]{0.70\columnwidth}
         \centering
         \includegraphics[width=\linewidth]{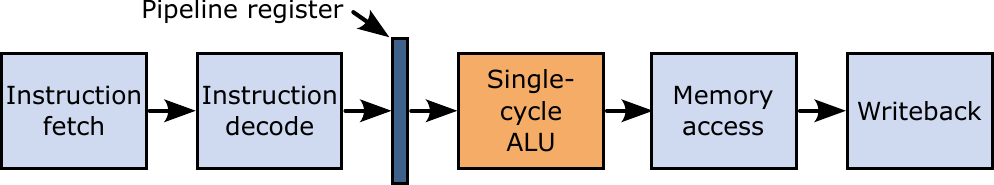}
         \caption{2-stage pipelined architecture.}
         \label{fig:microarch-two-stage}
     \end{subfigure}
     
     \begin{subfigure}[b]{0.8\columnwidth}
         \centering
         \includegraphics[width=\linewidth]{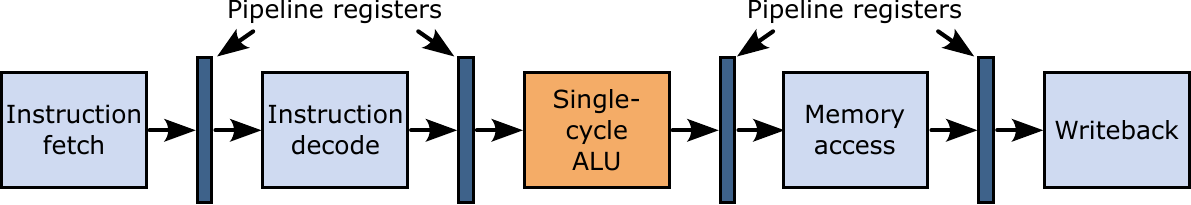}
         \caption{5-stage pipelined architecture.}
         \label{fig:microarch-five-stage}
     \end{subfigure}
    
    \caption{Ultra-low-power processor microarchitecture design options. {\it SERV, QERV, and PicoRV32 are multi-cycle architectures (a),  Ibex and Vex-2 are 2-stage pipelined architectures (b), and Vex-5 and Rocket are 5-stage pipelined architectures (c).}}
    \label{fig:microarch}
\end{figure}

We evaluate the SERV~\cite{serv:GITHUB}, QERV~\cite{serv:GITHUB}, PicoRV32~\cite{picorv32:GITHUB}, Ibex~\cite{ibex:GITHUB}, Vex~\cite{vexriscv:GITHUB}, and Rocket~\cite{rocket} 32-bit RISC-V cores, and we will now explain their microarchitectures in more detail. In all of these cores, instruction execution can be broken down into five key steps which the core must be able to perform to execute all RISC-V instructions. The core (1) fetches the instruction from memory, (2) decodes the instruction and retrieves its operands from the register file, and (3) performs the ALU-operation(s) the instruction requires. For load and store instructions, the ALU is used to compute the memory address, which in turn enables accessing memory (4).
In step (5), the core writes the output of the instruction to the register file (if the instruction produces output).
While all the cores we consider perform these five steps, \autoref{fig:microarch} illustrates that the manner in which they implement them varies widely. 

SERV, QERV, and PicoRV32 are all multi-cycle architectures, which means that an instruction is executed to completion over multiple clock cycles before the execution of the next instruction is initiated, i.e., there is only a single instruction in the datapath at any given time (see \autoref{fig:microarch-multi-cycle}).
This simplifies the control unit as the causes for hazards are significantly reduced and eliminates the need for forwarding logic, resulting in reduced area at the cost of lower performance.
SERV and QERV reduce area further by implementing an iterative ALU.
SERV takes this to the extreme by implementing a bit-serial ALU where a single bit of the final result is produced per cycle, resulting in a 32-cycle ALU latency. 
QERV trades latency against area by implementing a 4-bit wide (or quaternary) ALU, resulting in an 8-cycle ALU latency.
In comparison, the fully bit-parallel ALU of the PicoRV32 has a single-cycle latency.

The key alternative to a multi-cycle architecture is a pipelined architecture, which exploits instruction-level parallelism to execute parts of instructions in parallel. \autoref{fig:microarch-two-stage} and \ref{fig:microarch-five-stage} illustrate that this is enabled by separating (some of) the datapath units by pipeline registers.
Ibex and Vex-2 have a two-stage pipeline (see \autoref{fig:microarch-two-stage}), while Vex-5 and Rocket both have 5-stage pipelines (see \autoref{fig:microarch-five-stage}).
The pipeline registers enable Vex-2 and Ibex to fetch and decode one instruction in parallel with executing, accessing memory, and writing back the results of a different instruction. Vex-5 and Rocket on the other hand can work on up to five instructions in parallel.
At nominal supply voltage, deepening the pipeline makes it possible to increase the clock frequency and thereby performance --- because the amount of work that needs to be completed in a single clock cycle decreases when the number of stages increases --- but this comes at the cost of increased area and complexity --- because dependencies between instructions result in hazards that the core must handle such that the application executes correctly.

%% file: experiment.tex
\section{Experimental Setup}
\label{sec:experimental-setup}

We use Cadence Genus v19.15~\cite{genus} and Innovus v20.15~\cite{innovus} in a physical design flow (Verilog to GDSII) using our custom cell library.
The library is functional from 250\,mV, which was chosen as the target voltage for synthesis and physical flow of all the cores.  Each core is given a square floorplan with an area resulting
in around 70\% utilization after routing.  We constrain synthesis and place and route to the smallest clock period resulting in no timing violations, since our evaluation has shown that this results in improved energy efficiency.

We evaluate all the cores mentioned in \autoref{sec:cores}.  
Each core is configured with the RV32E profile intended for low-power embedded systems, which reduces the number of registers to 16.
They are configured for the smallest area by disabling as many features as possible, e.g., no hardware support for multiplication.
For the case of SERV, QERV, PicoRV32, and Ibex, a custom latch-based register file is used to further reduce their size.
Modifying the Vex and Rocket designs to use the latch-based register file proved challenging, and we instead used their respective conventional flip-flop-based register files.
Since our main focus is on the impact of the core implementation itself, we have not included a memory in the designs and assume a memory latency of a single cycle.

\begin{table}[t]
  \centering
  \caption{Results from final layout.}
  \label{tab:PRcores}
  \begin{tabular}{l|r|c|r}
    Core     & Clock Period  & Area                       & Transistors \\
    \hline \hline
    SERV     &   478\,ns     & 0.096\,mm\textsuperscript{2} &  25 648 \\
    QERV     &   512\,ns     & 0.109\,mm\textsuperscript{2} &  28 968 \\
    PicoRV32 &   686\,ns     & 0.235\,mm\textsuperscript{2} &  61 858 \\
    Vex-2    & 1 458\,ns     & 0.250\,mm\textsuperscript{2} &  62 294 \\
    Ibex     & 1 450\,ns     & 0.384\,mm\textsuperscript{2} & 103 630 \\
    Vex-5    &   998\,ns     & 0.423\,mm\textsuperscript{2} & 114 532 \\
    Rocket   & 1 434\,ns     & 0.792\,mm\textsuperscript{2} & 209 092 \\
    \hline
    Vex-2 (1.2\,V)  & 16\,ns & 0.090\,mm\textsuperscript{2} &  48 091 \\
    Vex-5 (1.2\,V) &  4\,ns & 0.160\,mm\textsuperscript{2} &  88 402 \\
  \end{tabular}
\end{table}

We used the SERV and PicoRV32 cores to identify the most energy efficient supply voltage for our cell library by comparing energy usage on a range of voltages from 250\,mV to 600\,mV.  Simulations show that both cores are most energy efficient with a supply voltage close to 300\,mV. 
All cores in this paper are, therefore, evaluated and compared with a supply voltage of 300\,mV. 
This is in line with earlier findings by other researchers~\cite{liu1993trading, wang2006sub}; the energy minimum typically lies below the absolute values of the inherent threshold voltages.

A summary of the subthreshold designs after performing the full physical flow is shown in the top part of \autoref{tab:PRcores}.
We also synthesized and evaluated the two Vex configurations using the conventional standard cell library designed for a nominal supply voltage of 1.2\,V.
We restrict the available cells to the same type and number of cells available in our own cell library.
This enables us to compare different architectural design choices and their impact in a conventional cell library and one optimized for subthreshold operation.
The results from the 1.2\,V cores are shown in the lower part of \autoref{tab:PRcores}.  Synopsys PrimeTime~\cite{primetime} 
was used for timing analysis.  

We selected eight benchmarks from MachSuite~\cite{machsuite:IISWC2014} as representative workloads for low-power embedded devices.
We used Siemens QuestaSim~\cite{questa}
to simulate each benchmark for each core using the netlist from the final layout.
The resulting waveforms were fed together with the netlist and parasitic information to Synopsys PrimeTime Power~\cite{primetime}
to estimate the average power and energy usage for each benchmark.
The estimated power for a small test case has been compared against an XCelium simulation using the same setup as explained in \autoref{sec:cell_lib_verification}.
This was done for both SERV and PicoRV32.  Energy from PrimeTime and from the XCelium simulations were found to be within 12\% of each other.

%% file: results.tex
\section{Results}

\begin{table}[tp]
  \centering
  \caption{Averages across all benchmarks.}
  \begin{tabular}{l|r|r|r|r}
    Core           & CPI   & Runtime & Power             & EPI \\
    \hline
    \hline
    SERV           & 57.94 & 14.68   &  2.85\,\textmu W  & 78.74\,pJ \\
    QERV           & 18.16 & 4.93    &  3.46\,\textmu W  & 32.18\,pJ \\
    PicoRV32       &  5.36 & 1.95    &  5.37\,\textmu W  & 19.74\,pJ \\
    Vex-2          &  1.29 & 1.00    &  3.73\,\textmu W  & 7.04\,pJ \\
    Ibex           &  1.59 & 1.22    &  6.13\,\textmu W  & 14.10\,pJ \\
    Vex-5          &  1.72 & 0.91    &  9.17\,\textmu W  & 15.76\,pJ \\
    Rocket         &  1.71 & 1.30    & 10.08\,\textmu W  & 24.72\,pJ \\
    \hline
    Vex-2 (1.2\,V) &  1.29 & 0.0112  &  2 324\,\textmu W & 49.20\,pJ \\
    Vex-5 (1.2\,V) &  1.72 & 0.0037  & 17 671\,\textmu W & 124.78\,pJ \\
  \end{tabular}
  \label{tab:results_summary}
\end{table}

A summary of the results is found in \autoref{tab:results_summary}; runtime is normalized to Vex-2.  Vex-2 is clearly the most energy efficient core, having an EPI of only 7\,pJ.  SERV is the most power efficient core with an average power of 2.85\,\textmu W, while Vex-5 is the fastest among the subthreshold cores.

\begin{figure*}[t]
  \centering
\begin{subfigure}{0.32\textwidth}
  \centering
  \includegraphics[width=\textwidth]{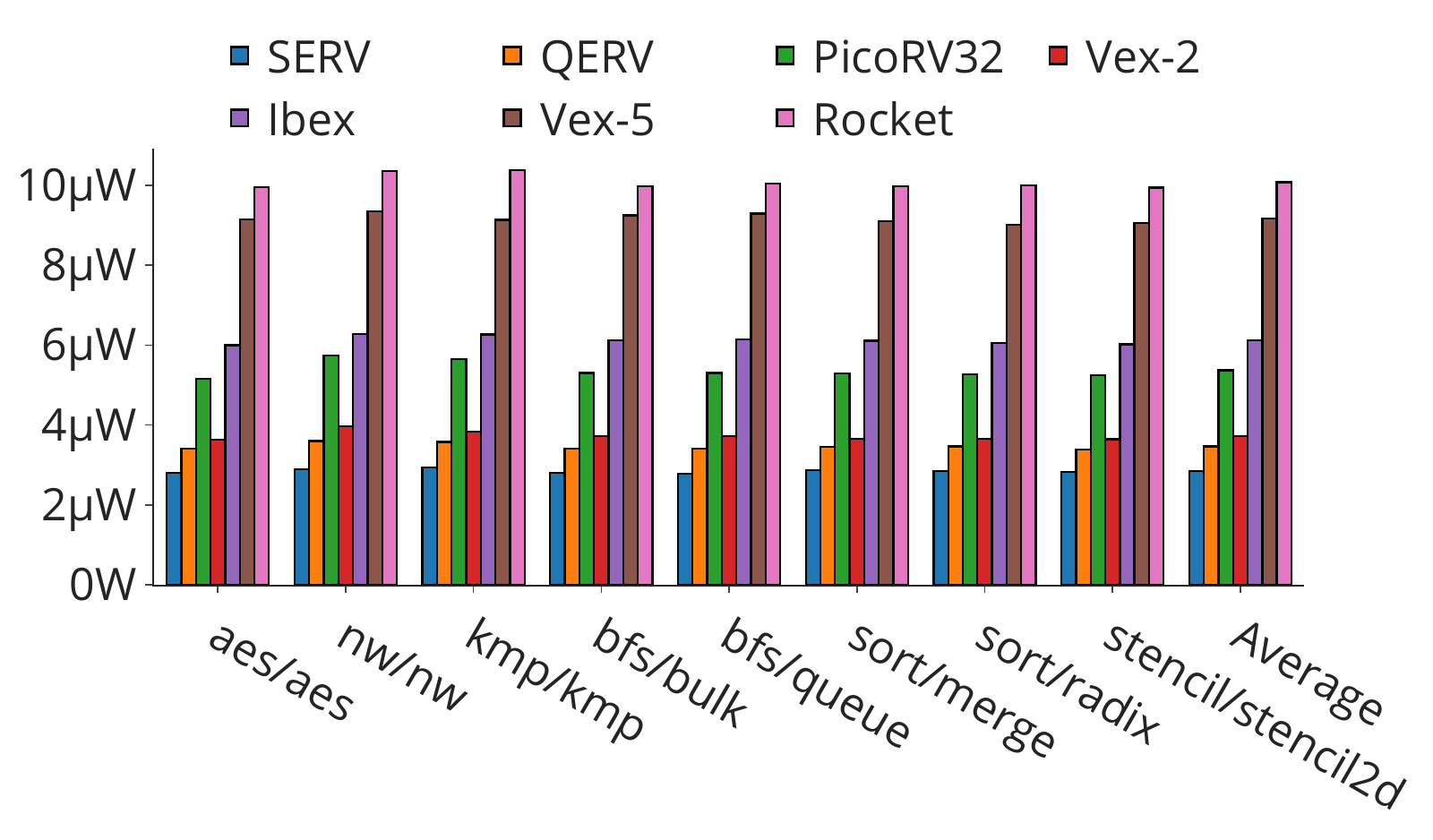}
  \caption{Average power (subthreshold cores).}
  \label{fig:power_subvth}
\end{subfigure}
\hfill
\begin{subfigure}{0.32\textwidth}
  \centering
  \includegraphics[width=\textwidth]{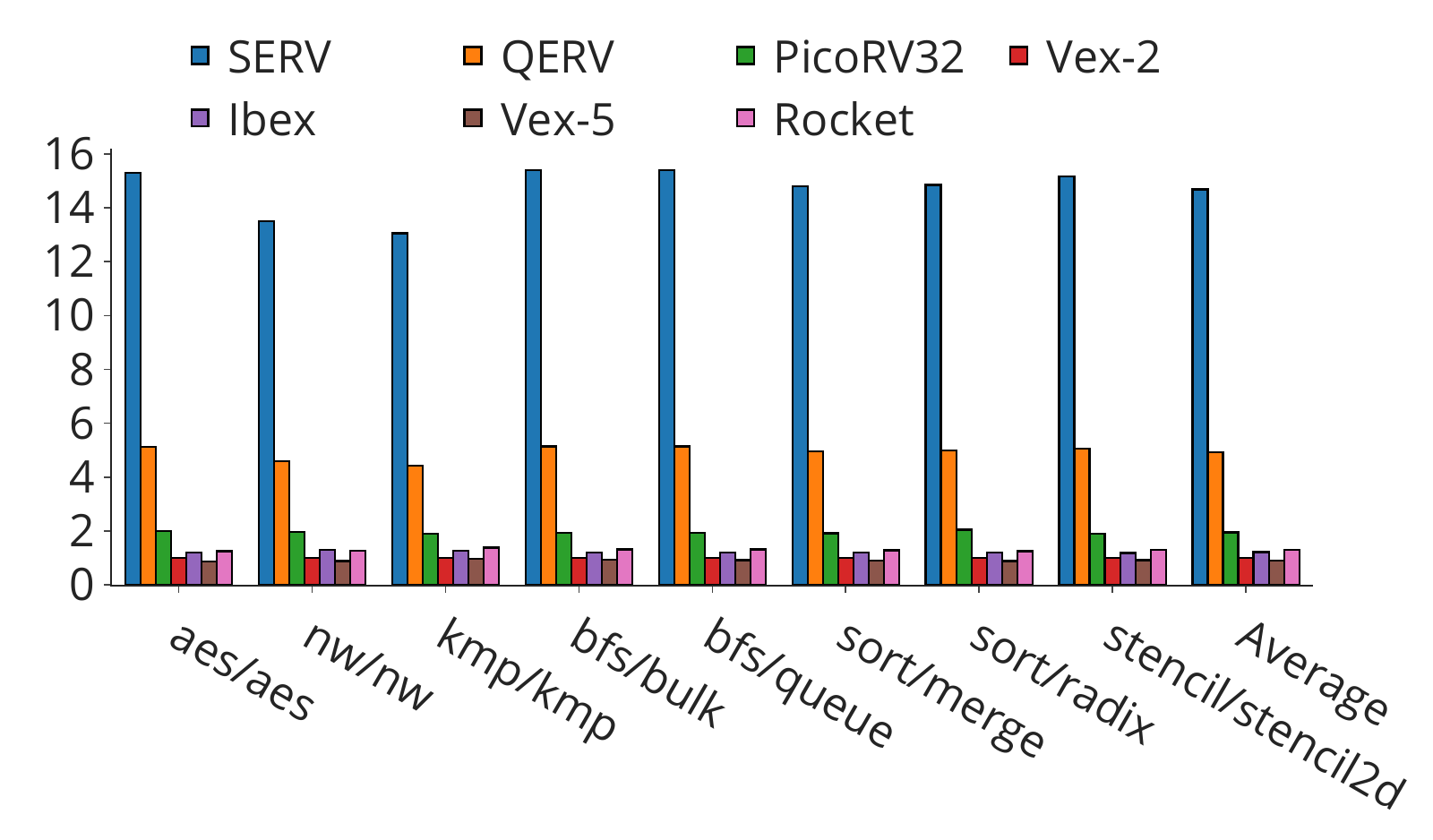}
  \caption{Execution time (subthreshold cores).}
  \label{fig:runtime}
\end{subfigure}
\hfill
\begin{subfigure}{0.32\textwidth}
  \centering
  \includegraphics[width=\textwidth]{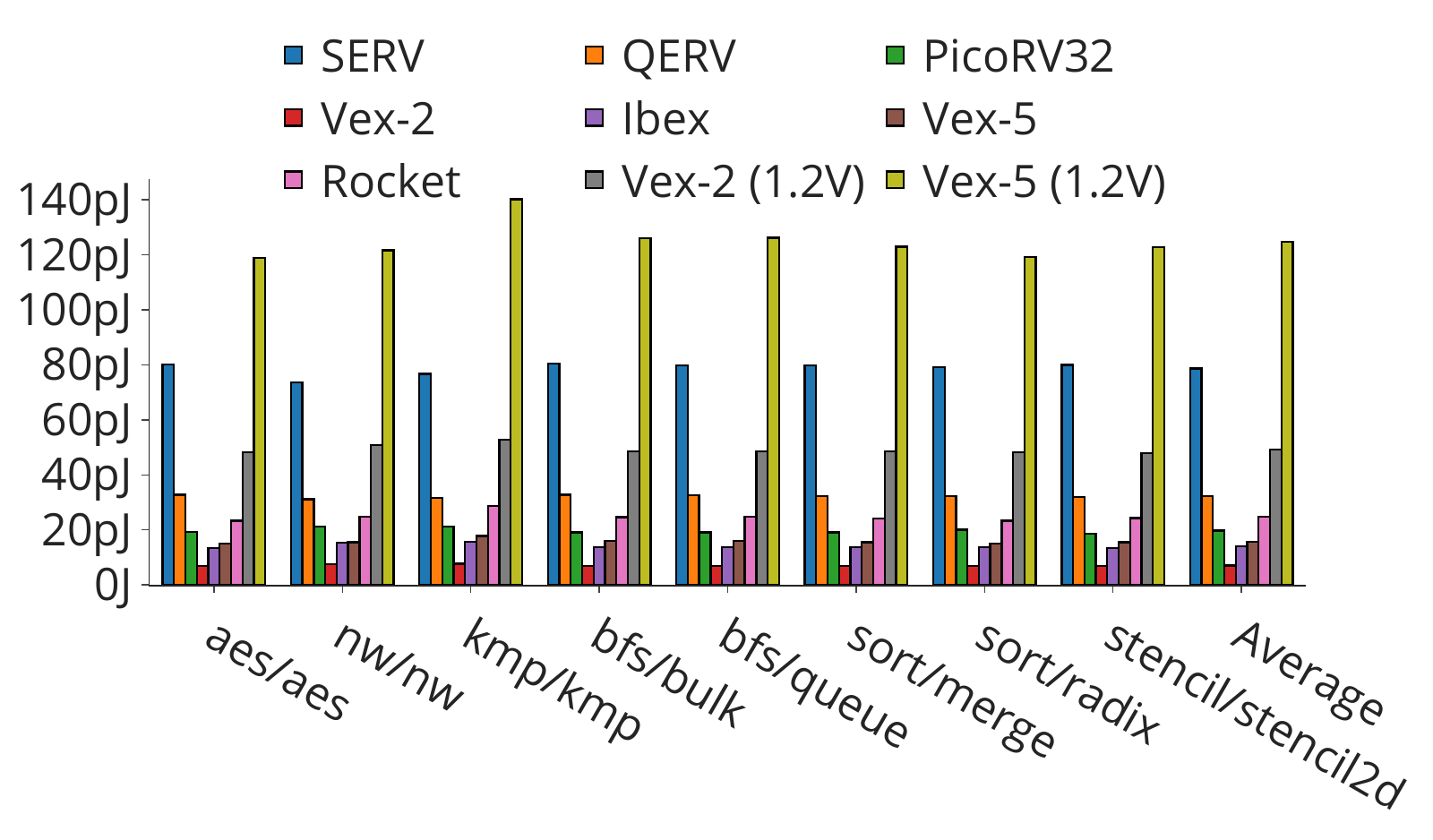}
  \caption{EPI (all cores).}
  \label{fig:epi}
\end{subfigure}
  \caption{Power consumption, execution time, and Energy Per Instruction (EPI) across benchmarks and cores.}
  \label{fig:power_runtime_epi}
\end{figure*}

\begin{figure*}[t]
  \centering
\begin{subfigure}{0.32\textwidth}
  \centering
  \includegraphics[width=\textwidth]{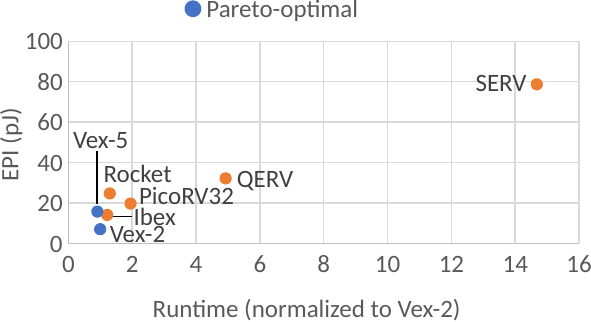}
  \caption{Runtime versus energy efficiency.}
  \label{fig:runtim-vs-epi}
\end{subfigure}
\hfill
\begin{subfigure}{0.32\textwidth}
  \centering
  \includegraphics[width=\textwidth]{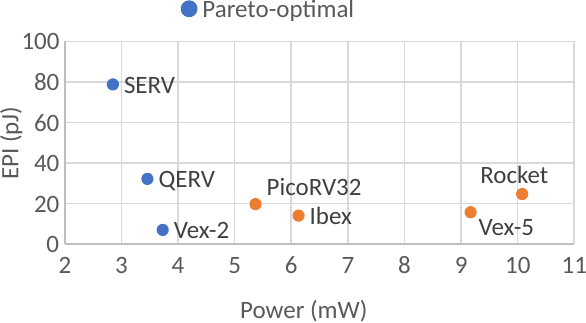}
  \caption{Power versus energy efficiency.}
  \label{fig:power-vs-epi}
\end{subfigure}
\hfill
\begin{subfigure}{0.32\textwidth}
  \centering
  \includegraphics[width=\textwidth]{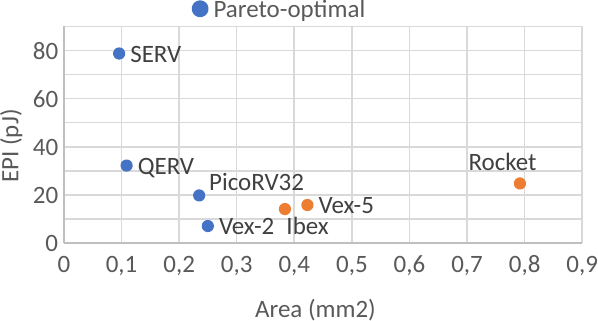}
  \caption{Area versus energy efficiency.}
  \label{fig:area-vs-epi}
\end{subfigure}
  \caption{Pareto-optimal designs when optimizing for energy efficiency (EPI) and runtime (a), power (b), or area (c).}
  \label{fig:pareto}
\end{figure*}

\subsection{Power, Performance, and Energy Efficiency}
\label{subsec:power-perf-energy}

\noindent
\textbf{Power.}
\autoref{fig:power_subvth} shows power for the subthreshold cores.  The horizontal axis shows benchmark runs for each core, grouped by benchmark.  Cores are ordered by their size in number of transistors.  The vertical axis is average power.
As can be seen, power is quite constant across the different benchmarks.  Power is also correlated with area, so the smallest core, SERV, is also the most power efficient.
PicoRV32 has almost the same power consumption as the two-stage Ibex, while Vex-2 has a significantly lower power consumption than both PicoRV32 and Ibex.
Deeper pipelines clearly result in higher power consumption, as shown for the 5-stage Vex-5 and Rocket cores.

\autoref{tab:results_summary} reports average power consumption over all the benchmarks for both the subthreshold cores and the 1.2\,V cores; The 1.2\,V cores are three orders of magnitude more power hungry than the subthreshold cores (mW instead of $\mu$W).

\noindent
\textbf{Performance.}
Execution time is shown in \autoref{fig:runtime}.  The horizontal axis shows benchmark runs for each core and is ordered the same way as in \autoref{fig:power_subvth}.  The vertical axis is execution time, normalized to Vex-2.  
There are significant differences, with SERV being 15$\times$ slower than Vex-2.
QERV has a 4-bit wide ALU and achieves 3$\times$ better performance than the bit-serial SERV.
In general, all the multi-cycle cores are slow because of their low average CPI (cycles per instruction).
Average CPI is shown in \autoref{tab:results_summary}.

The pipelined cores have longer clock periods than the multi-cycle cores (see \autoref{tab:PRcores}).
Nevertheless, the pipelined cores are much faster than the multi-cycle cores because they all have a CPI close to one. At 300\,mV, Vex-2 improves performance by 15$\times$ and 5$\times$ compared to SERV and QERV, respectively, whereas Ibex yields performance improvements of 12$\times$ and 4$\times$.  Even though both Vex-2 and Ibex are 2-stage cores, they differ in their CPI.  This is due to Ibex stalling one cycle for every load and store instruction~\cite{ibex_pipeline}, resulting in a 23\% increase in average CPI compared to Vex-2. 
Vex-5 is the fastest core because of a relatively low CPI and the lowest clock period, and is on average 9\% faster than Vex-2.  Vex-2 is the second-fastest core and has the lowest CPI.

Table~\ref{tab:results_summary} shows that the 1.2\,V cores are two orders of magnitude faster than the subthreshold cores, resulting in Vex-5 (1.2\,V) being the overall fastest core. 

\begin{table*}[tp]
  \centering
  \caption{Instance count and timing for the critical path.}
  \resizebox{\textwidth}{!}{
  \begin{tabular}{l|r|r|r|r|r|r|r|r|r|r|r|r|r|r}
                     & \multicolumn{2}{c|}{Total} & \multicolumn{3}{c|}{Buffers} & \multicolumn{3}{c|}{Inverters} & \multicolumn{3}{c|}{Comb.} & \multicolumn{3}{c}{Flip-flops} \\
                     & Count & Delay & Count & \multicolumn{2}{c|}{Delay} & Count & \multicolumn{2}{c|}{Delay} & Count & \multicolumn{2}{c|}{Delay} & Count & \multicolumn{2}{c}{Delay} \\
    \hline
    \hline
    Vex-2	&91	&1398,75	&22	&447,36	&31,98\%	&21	&240,08	&17,16\%	&47	&668,64	&47,80\%	&1	&42,67	&3,05\%\\
    Vex-5	&47	&968,1	&31	&705,36	&72,86\%	&6	&95,1	&9,82\%	&9	&115,53	&11,93\%	&1	&52,11	&5,38\%\\
    Vex-2 (1.2\,V)	&90	&16,14	&1	&0,19	&1,18\%	&5	&0,72	&4,46\%	&83	&15,04	&93,18\%	&1	&0,19	&1,18\%\\
    Vex-5 (1.2\,V)	&33	&3,86	&0	&0	&0,00\%	&5	&0,61	&15,80\%	&27	&3,08	&79,79\%	&1	&0,17	&4,40\%\\
  \end{tabular}
  }
  \label{tab:critical_path}
\end{table*}

\begin{table}[tp]
  \centering
  \caption{Area breakdown.}
  \begin{tabular}{l|r|r|r|r}
    Core         & Buffers & Inverters & Comb.  & Flip-flops   \\
    \hline
    \hline
    Vex-2        & 31.5\%  & 11.5\%    & 33.8\% & 23.2\% \\
    Vex-5        & 34.5\%  & 10.0\%    & 29.5\% & 26.0\% \\
    \hline
    Vex-2 (1.2\,V) &  0.1\%  &  8.7\%    & 56.3\% & 34.4\% \\
    Vex-5 (1.2\,V) &  0.0\%  &  8.1\%    & 52.7\% & 38.7\% \\
  \end{tabular}
  \label{tab:area}
\end{table}

\noindent
\textbf{Energy per instruction.}
Our energy efficiency metric is energy per instruction (EPI), and \autoref{fig:epi} shows EPI for all benchmarks.  The horizontal axis shows benchmark runs for each core, ordered the same way as in Figures \ref{fig:power_subvth} and \ref{fig:runtime}.  The vertical axis is EPI, where a core with a lower number means a more energy efficient core.

Vex-2 is clearly the most energy efficient core for all benchmarks, with an average EPI of 7\,pJ.  This is both a result of its low power consumption, seen in \autoref{fig:power_subvth}, and its low execution time, seen in \autoref{fig:runtime}.  SERV is the least energy efficient core (excluding the 1.2\,V cores) with an EPI of 79\,pJ, which is 11$\times$ larger than the EPI of Vex-2.  Even though SERV is the core with the lowest power consumption (\autoref{fig:power_subvth}), its high execution time (\autoref{fig:runtime}) makes it more energy-hungry than all the other subthreshold cores. 
At 1.2\,V, Vex-2 and Vex-5 have 7--8$\times$ the EPI of their subthreshold equivalents, i.e., they are roughly one order of magnitude less energy efficient. This observation is in line with prior work~\cite{6359800,Beiu}. The massive difference in energy efficiency results in Vex-5 (1.2\,V) being the least energy efficient core in our evaluation (EPI of 125\,pJ).

\noindent
\textbf{Pareto-optimal design points.}
When designing for ultra-energy-efficient devices, multiple design constraints often have to be balanced against each other.
Such devices are commonly mass-produced, and low price (small area) is usually a competitive advantage.  They are also often powered by energy harvesting, which enforce power consumption limitations, while still requiring to be performant.
\autoref{fig:pareto} shows the evaluated cores when energy efficiency (EPI) is plotted against runtime~(a), power~(b), and area~(c).
The figures show that Vex-2 is a Pareto-optimal design for all three cases, with Vex-5 being Pareto optimal if a faster core is required (9\%), but at a 2.2$\times$ reduction in energy efficiency and 1.7$\times$ increase in area.
For power and area constrained devices, we find that SERV and QERV are Pareto optimal.
Their bit/quaternary-serial operation reduces their area by more than 2$\times$ compared to Vex-2, which also results in lower power consumption at 24\% and 6\%, respectively.
Their main drawback is the increased runtime at 15$\times$ and 5$\times$, which results in lower energy efficiency at 11$\times$ and 5$\times$ compared to Vex-2, respectively.
PicoRV32 is Pareto optimal in terms of area, with a 6\% decrease at a cost of 2.8$\times$ reduction in energy efficiency compared to Vex-2.
Ibex and Rocket are found to \emph{not} be Pareto optimal.

\subsection{Pipelining for Ultra-Low-Power Cores}

The Vex-2 and Vex-5 cores are constructed from the same source code, only with different configurations.
This makes them suitable for studying how the number of pipeline stages affect runtime, power, energy, and area.
Pipelining improves execution time by splitting the critical path with a pipeline register, thus increasing clock frequency. 
Extra registers and pipelining logic leads to an increase in area of approximately 1.7$\times$ when going from a 2 to 5 stages. 
As shown in \autoref{tab:PRcores}, at 1.2\,V we achieve a clock frequency that is almost 4$\times$ faster when going from a 2-stage to a 5-stage pipeline.
In subthreshold, the clock frequency is only 1.5$\times$ faster.
As the CPI also increases when going to five stages, there is only a 9\% improvement in execution time when using a 5-stage pipeline in subthreshold, which makes pipelining much less attractive in subthreshold compared to 1.2\,V. 

In order to understand why pipelining the subthreshold Vex does not experience the same benefits as the standard supply voltage Vex, we investigated the critical paths of the different Vex cores.  \autoref{tab:critical_path} shows which cells constitute the critical paths of the different cores. 
We see that for Vex-2 in subthreshold, buffers contribute to 32\% of the delay of the critical path.
Only 48\% of the delay is due to combinatorial cells, i.e., the cells that implement the logic functions.
For Vex-2 at 1.2\,V, the buffers only contribute to 1.2\% of the total delay, while combinatorial cells contribute 93\%.
This demonstrates the main difference between the subthreshold cores and the 1.2\,V cores: The subthreshold cores have much higher buffer requirements.  Due to its low current levels, subthreshold circuits are much more penalized by long wires that load the subthreshold gates and increase delays substantially.  This forces the EDA flow to insert buffers to meet timing requirements.  Relaxing the timing requirements would get rid of the buffers, but the clock frequency would then drop to such a degree that energy efficiency would decrease.  This changes the effect pipelining has in subthreshold. 
Comparing Vex-5 to Vex-2 in \autoref{tab:critical_path} shows that pipelining to a large degree reduces the number of combinatorial gates, but not the buffers.  The critical path of Vex-5 in subthreshold has 73\% of its total delay spent on buffers, while only 12\% is spent on combinatorial cells.  At 1.2\,V the situation is very different; there are no buffers in the critical path. 
Increased buffering is seen all over the core.  \autoref{tab:area} shows which cells are contributing to the area of the cores.  In subthreshold, buffers take up 31.5\% of the area of the Vex-2 core, while they are almost nonexistent in the 1.2\,V core.  When introducing pipelining to the subthreshold core, the proportion of buffers increases to 34.5\% because of the larger area requiring longer wires. 

When looking at energy efficiency, deeper pipelines are even less attractive.
Evident from \autoref{fig:epi} and \autoref{tab:results_summary}, the 5-stage Vex core is less energy efficient than the 2-stage Vex core.
Even though the clock period (and therefore also the execution time) decreases for the 5-stage core, the power increases to such a degree that the 5-stage core ends up being much less energy efficient.
This is the case for both subthreshold and nominal voltages.
The power increases 7.6$\times$ at 1.2\,V and 2.5$\times$ at subthreshold.
Some of this increase in power is due to larger area, and some of it is due to higher switching activity.
Leakage is a major component of the average power in subthreshold, but almost insignificant at nominal voltage.
This means that the power of the subthreshold 5-stage core is less affected by high switching activity than for nominal voltages, and the power and energy penalty is not as high.
However, as the improvement in execution time is so slight, there are seemingly no good reasons for using a 5-stage pipeline RISC-V core when designing for subthreshold operation.  

%% file: conclusion.tex
\section{Conclusion}

We have studied several open-source RISC-V cores to understand how to design energy efficient standard-ISA processors for subthreshold operation.
With power simulations, we found that 2-stage pipelined Vex-2 is the most energy efficient of the cores we evaluate.
All the multi-cycle cores perform significantly worse than the pipelined cores due to their high average CPI, which negatively affect execution times and energy. 
Modifying the pipeline from two to five stages, as implemented by Vex-5, was found to have limited effect on execution time in subthreshold.  A significant amount of buffers are necessary to keep timing, and the buffering needs increase when going to five stages.  These buffers lead to limited reduction in cycle times when adding pipeline stages, but the area and power increases significantly. The 5-stage pipelines are, therefore, much less energy efficient than the 2-stage pipelines, and we conclude that the 2-stage pipeline is the sweet spot for  energy efficient subthreshold RISC-V cores.